\documentclass[onecolumn,prb,floatfix,nofootinbib,preprint]{revtex4}

\usepackage{graphicx}
\usepackage{amsmath,bm}
\usepackage{slashed}
\usepackage{multirow}
\usepackage{setspace}
\usepackage{boxedminipage}

\begin{document}

\begin{center}
{\large \bf Renormalization group running of neutrino parameters}

\vspace{0.3cm}

{\sc Tommy Ohlsson}~$^{a,}$~\footnote{email: tohlsson@kth.se}
~and~  {\sc Shun Zhou}~$^{a,~b,}$~\footnote{email: zhoush@ihep.ac.cn}

\vspace{0.2cm}

{$^a$~Department of Theoretical Physics, School of Engineering Sciences, \\
KTH Royal Institute of Technology, AlbaNova University Center, \\
106 91 Stockholm, Sweden} \\
{$^b$~Theoretical Physics Division, Institute of High Energy Physics, Chinese Academy of Sciences, Beijing 100049, China}
\end{center}

\noindent {\sl Neutrinos are the most elusive particles in our universe. They have masses at least one million times smaller than the electron mass, carry no electric charge, and very weakly interact with other particles, meaning they are rarely captured in terrestrial detectors. Tremendous efforts in the past two decades have revealed that neutrinos can transform from one type to another as a consequence of neutrino oscillations---a quantum mechanical effect over macroscopic distances---yet the origin of neutrino masses remains puzzling. The physical evolution of neutrino parameters with respect to energy scale may help elucidate the mechanism for their mass generation.}

\vspace{1cm}

\noindent Ever since their discovery in the 1950s (ref.~\onlinecite{Cowan:1992xc}), neutrinos have continued to surprise us. In the Standard Model (SM) of elementary particle physics, neutrinos are massless particles. However, since the results from the Super-Kamiokande experiment in 1998 (ref.~\onlinecite{Fukuda:1998mi}), the phenomenon of neutrino oscillations has been well established, indicating that neutrinos do have nonzero and non-degenerate masses and that they can convert from one flavour to another~\cite{PDG:2012}. This important result was followed by a boom of results from several international collaborations. Certainly, these results have pinned down the values of the various neutrino parameters to an incredible precision, especially considering that neutrinos are extremely elusive particles and the corresponding experiments are extraordinarily complex~\cite{Ohlsson:2012n}. Currently operating experiments and future investigations under construction are aimed at determining the missing neutrino parameters, such as the CP-violating phase (which can be important for understanding the matter--antimatter asymmetry in the universe), the sign of the large mass-squared difference for neutrinos, and the absolute neutrino mass scale. In addition, the cubic-kilometer-scale neutrino telescope at the South Pole, IceCube~\cite{Ahrens:2002dv}, has been successfully constructed to search for ultrahigh-energy astrophysical neutrinos, while a number of underground experiments are looking for neutrinoless double beta decay (see refs~\onlinecite{Cartlidge:2012,Agostini:2013,Albert:2014,Auger:2012,Gando:2013}) and others are waiting for neutrino bursts from galactic supernova explosions (see refs~\onlinecite{Antoniolli:2004,Scholberg:2012}).

However, the origin of neutrino masses and lepton flavour mixing remains a mystery, and calls for new physics beyond the SM. It is believed that new physics should appear somewhere above the electroweak scale (that is, $\Lambda^{}_{\rm EW} \sim 10^2~{\rm GeV}$) but below the Planck scale (that is, $\Lambda^{}_{\rm P} \sim 10^{19}~{\rm GeV}$) for the following reasons. First, the smallness of neutrino masses can be ascribed to the existence of superheavy particles, whose masses are close to the grand-unified-theory (GUT) scale (for example, $\Lambda^{}_{\rm GUT} \sim 10^{16}~{\rm GeV}$), such as right-handed neutrinos in the canonical seesaw models~\cite{Minkowski:1977sc,Yanagida:1979ss,Gell-Mann:1979ss,Glashow:1979ss,Mohapatra:1979ia}. Moreover, the out-of-equilibrium and CP-violating decays of heavy right-handed neutrinos in the early universe can produce a lepton number asymmetry, which will be further converted into a baryon number asymmetry~\cite{Fukugita:1986hr}. Therefore, the canonical seesaw mechanism combined with so-called leptogenesis provides an elegant solution to the generation of tiny neutrino masses and the matter--antimatter asymmetry in our universe. Second, the strong hierarchy in charged-fermion masses (that is, $m^{}_t \gg m^{}_c \gg m^{}_u$, $m^{}_b \gg m^{}_s \gg m^{}_d$, and $m^{}_\tau \gg m^{}_\mu \gg m^{}_e$) and the significant difference between quark and lepton mixing patterns (that is, three small quark mixing angles while two large and one small leptonic mixing angles) could find their solutions in the framework of grand-unified theories extended by a flavour symmetry~\cite{Altarelli:2010gt,King:2013eh}. Therefore, an attractive and successful flavour model usually works at a superhigh-energy scale, where quarks and leptons are unified into the same multiplets of the gauge group but assigned into different representations of the flavour symmetry group. Third, the SM Higgs particle with a mass of $126~{\rm GeV}$ has recently been discovered at the Large Hadron Collider at CERN in Geneva, Switzerland~\cite{Aad:2012tfa,Chatrchyan:2012ufa}. If this is further confirmed by future precision measurements and the top-quark mass happens to be large, the SM vacuum will become unstable around the energy scale $10^{12}~{\rm GeV}$. In this case, new physics has to show up to stabilize the SM vacuum~\cite{Buttazzo:2013uya}. In the canonical seesaw model with heavy right-handed neutrinos, the SM vacuum is actually further destabilized. However, if an extra scalar singlet is introduced to generate right-handed neutrino masses, the SM vacuum can be stabilized and the tiny neutrino masses are explained via the seesaw mechanism~\cite{EliasMiro:2012ay}. Hence, the assumption that neutrino masses and lepton flavour mixing are governed by new physics at a superhigh-energy scale is well motivated.

The experimental results will guide us to the true theory of neutrino masses, lepton flavour mixing, and CP violation. At the same time, they will also rule out quite a large number of currently viable flavour models. However, this will only be possible if the renormalization group running of neutrino parameters, which describes their physical evolution with respect to energy scale, is properly taken into account. Thus, it may help to elucidate the mechanism for neutrino mass generation. The aim of this review is to examine neutrino renormalization group running in more detail. First, we briefly summarize the current status of neutrino parameters and the primary goals of future neutrino experiments, and present a general discussion about the effective theory approach and renormalization-group equations in particle physics. Then, we consider several typical neutrino mass models in the framework of supersymmetric and extra-dimensional theories, and the running behaviour of neutrino parameters is described and explained. Finally, the impact of renormalization-group running on flavour model building and leptogenesis is illustrated and emphasized.

\section*{Neutrino parameters at low energies}
\label{sec:neu_para}

\noindent Neutrinos are produced in beta decay of radioactive nuclei, nuclear fusion in the Sun, collisions between nucleons in the earth atmosphere and cosmic-ray particles, and in the man-made high-energy accelerators. Since they are always accompanied by the charged leptons $e$, $\mu$, and $\tau$ in production, it is convenient to define the neutrino flavour eigenstates $\{|\nu^{}_e\rangle, |\nu^{}_\mu\rangle, |\nu^{}_\tau\rangle\}$ and discriminate them according to the corresponding charged leptons. The neutrino flavour eigenstates $|\nu^{}_\alpha\rangle$ (for $\alpha = e, \mu, \tau$) are related to three neutrino mass eigenstates $\{|\nu^{}_1\rangle, |\nu^{}_2\rangle, |\nu^{}_3\rangle\}$ with definite masses $\{m^{}_1, m^{}_2, m^{}_3\}$ by the superposition $|\nu^{}_\alpha\rangle = U^{}_{\alpha 1} |\nu^{}_1\rangle + U^{}_{\alpha 2} |\nu^{}_2\rangle + U^{}_{\alpha 3} |\nu^{}_3\rangle$, where the $3\times 3$ unitary matrix $U$ is the so-called lepton flavour mixing matrix~\cite{Pontecorvo:1957cp,Maki:1962mu,Pontecorvo:1967fh}. It is conventional to parameterize $U$ by three Euler-like mixing angles $\{\theta^{}_{12}, \theta^{}_{13}, \theta^{}_{23}\}$ and three CP-violating phases $\{\delta, \rho, \sigma\}$, namely~\cite{PDG:2012}
\begin{eqnarray}
U = \left(\begin{matrix}1 & 0 & 0 \cr 0 & c_{23} & s_{23} \cr 0 & -s_{23} & c_{23}\end{matrix}\right) \left(\begin{matrix}c_{13} & 0 & s_{13} e^{-i\delta} \cr 0 & 1 & 0 \cr -s_{13} e^{i\delta} & 0 & c_{13}\end{matrix}\right) \left(\begin{matrix}c_{12} & s_{12} & 0 \cr -s_{21} & c_{12} & 0 \cr 0 & 0 & 1\end{matrix}\right) \left(\begin{matrix}e^{i\rho} & 0 & 0 \cr 0 & e^{i\sigma} & 0 \cr 0 & 0 & 1\end{matrix}\right)
\label{eq:MNSP}
\end{eqnarray}
with $c^{}_{ij} \equiv \cos \theta^{}_{ij}$ and $s^{}_{ij} \equiv \sin \theta^{}_{ij}$ for $ij = 12, 13, 23$. As a consequence of quantum interference among the three neutrino mass eigenstates, neutrinos can transform from one flavour to another, when propagating from the sources to the detectors. This phenomenon of neutrino flavour oscillations will be absent if either the two independent neutrino mass-squared differences $\Delta m^2_{21} \equiv m^2_2 - m^2_1$ and $\Delta m^2_{31} \equiv m^2_3 - m^2_1$ (or $\Delta m^2_{32} \equiv m^2_3 - m^2_2$) or the three leptonic mixing angles $\{\theta^{}_{12}, \theta^{}_{13}, \theta^{}_{23}\}$ are vanishing. Note that we will use $\Delta m^2_{31}$ instead of $\Delta m^2_{32}$.

Thanks to a number of elegant experiments in the past two decades~\cite{PDG:2012}, the phenomenon of neutrino flavour oscillations has now been firmly established. The latest global analysis of data from all existing past and present neutrino oscillation experiments provides our best knowledge on the neutrino mixing parameters, as shown in Table~\ref{tab:parameter_values}. Note that $\Delta m^2_{31}$ has been used in ref.~\onlinecite{Forero:2014bxa} to fit the oscillation data in both cases of normal neutrino mass hierarchy (that is, $m^{}_1 < m^{}_2 < m^{}_3$) and inverted neutrino mass hierarchy (that is, $m^{}_3 < m^{}_1 < m^{}_2$), only the results from ref.~\onlinecite{Forero:2014bxa} are listed in this table in order to get a ballpark feeling of the values of the neutrino parameters. Two other independent global-fit analyses in refs~\onlinecite{Capozzi:2013csa} and \onlinecite{GonzalezGarcia:2012sz} yield different best-fit values. However, the $3\sigma$ confidence intervals of neutrino parameters from all three groups are indeed consistent.
\begin{table}[t]
\renewcommand{\arraystretch}{1.4}
\begin{center}
\vspace{-0.25cm} \caption{{\bf Latest measurements and global-fit results of neutrino parameters.}}
\label{tab:parameter_values}
\vspace{0.2cm}
\begin{tabular}{ccc}
Parameters &  ~Neutrino oscillation experiments~\footnote{The experiments that dominate the accuracy of particular neutrino parameters determination are shown.}~ & Global-fit results~\footnote{The best-fit values and $1\sigma$ uncertainties are taken from ref.~\onlinecite{Forero:2014bxa}.} \\
\hline
$\Delta m^2_{21}$ & KamLAND ($\overline{\nu}^{}_e \to \overline{\nu}^{}_e$)~\cite{Gando:2010aa} & $[7.60^{+0.19}_{-0.18}] \cdot 10^{-5}~{\rm eV}^2$ \\[5mm]
\multirow{2}{3cm}{\centering $\Delta m^2_{31}$} & T2K ($\nu^{}_\mu \to \nu^{}_\mu$)~\cite{Abe:2014ugx} & ~$+[2.48^{+0.05}_{-0.07}] \cdot 10^{-3}~{\rm eV}^2$ (NH) \\
~ & MINOS ($\overline{\nu}^{}_\mu \to \overline{\nu}^{}_\mu$, $\nu^{}_\mu \to \nu^{}_\mu$)~\cite{Adamson:2013whj} & ~$-[2.38^{+0.05}_{-0.06}] \cdot 10^{-3}~{\rm eV}^2$ (IH)~ \\[5mm]
\multirow{3}{3cm}{\centering $\theta^{}_{12}$} & solar neutrinos ($\nu^{}_e \to \nu^{}_e$) & \multirow{3}{4cm}{\centering ${34.63^\circ}^{+1.02^\circ}_{-0.98^\circ}$}  \\
~ & Borexino~\cite{Bellini:2011rx}, SNO~\cite{Aharmim:2008kc,Aharmim:2009gd}, & ~ \\
~ & Super-Kamionkande I-IV~\cite{Renshaw:2014awa} & ~ \\[5mm]
\multirow{2}{3cm}{\centering $\theta^{}_{13}$} & Daya Bay ($\overline{\nu}^{}_e \to \overline{\nu}^{}_e$)~\cite{An:2013zwz}& ${8.80^\circ}^{+0.37^\circ}_{-0.39^\circ}$ (NH)\\
~ & RENO ($\overline{\nu}^{}_e \to \overline{\nu}^{}_e$)~\cite{Ahn:2012nd} & ${8.91^\circ}^{+0.35^\circ}_{-0.36^\circ}$ (IH)~ \\[5mm]
\multirow{3}{3cm}{\centering $\theta^{}_{23}$} & atmospheric neutrinos & \multirow{2}{3cm}{\centering ${48.9^\circ}^{+1.6^\circ}_{-7.4^\circ}$ (NH)} \\
~ & ($\overline{\nu}^{}_\mu \to \overline{\nu}^{}_\mu$, $\nu^{}_\mu \to \nu^{}_\mu$) & \multirow{2}{3cm}{\centering ${49.2^\circ}^{+1.5^\circ}_{-2.5^\circ}$ (IH)~} \\
~ & Super-Kamiokande I-IV~\cite{Himmel:2013jva} & \\[5mm]
\multirow{2}{3cm}{\centering $\delta$} & \multirow{2}{3cm}{\centering --} & ~${241^\circ}^{+115^\circ}_{-68^\circ}$ (NH) \\
~ & ~ & ~${266^\circ}^{+62^\circ}_{-57^\circ}$ ~(IH)~ \\
\hline
\end{tabular}
\\[5mm]
{\footnotesize IH, inverted neutrino mass hierarchy; NH, normal neutrino mass hierarchy.}
\end{center}
\end{table}

At present, although there are weak hints for a nonzero Dirac CP-violating phase $\delta$ (see the last row of Table~\ref{tab:parameter_values}), it is fair to say that no direct and significant experimental constraints exist for the leptonic CP-violating phases. Furthermore, since neutrino oscillation experiments are blind to the Dirac or Majorana nature of neutrinos and to the Majorana CP-violating phases $\{\rho, \sigma\}$, it is still an open question whether neutrinos are Dirac or Majorana particles. In the latter case, neutrinos are their own antiparticles, which would lead to neutrinoless double-beta decay of some nuclear isotopes and can hopefully be confirmed with this kind of experiments~\cite{Rodejohann:2011mu}. The primary goals of ongoing and forthcoming neutrino oscillation experiments are to precisely measure the three leptonic mixing angles, to determine the neutrino mass hierarchy, and to discover the leptonic Dirac CP-violating phase. In addition, non-oscillation neutrino experiments aim to pin down the absolute neutrino masses and to probe the Majorana nature of neutrinos.

\section*{Confronting theories with experiments}
\label{sec:theo_exp}

\noindent Although most neutrino parameters have already been measured with a reasonably good precision, the origin of tiny neutrino masses and bi-large lepton flavour mixing remains elusive. In order to accommodate tiny neutrino masses, one may have to go beyond the SM at the electroweak scale and explore new physics at a superhigh-energy scale. In this case, an immediate question is how to compare theoretical predictions at a high-energy scale with the observables at a low-energy scale. With this question in mind, we present a brief account of effective theories and renormalization group running, and describe how neutrinos fit into this framework.

\vspace{5mm}

\noindent {\bf Effective theory approach.} The effective theory approach is very useful, and sometimes indispensable in particle physics, where interesting phenomena appear at various energy scales. The basic premise for this approach to work well is that the dynamics at low-energy scales (or large distances) does not depend on the details of the dynamics at high-energy scales (or short distances). For instance, the energy levels of a hydrogen atom are essentially determined by the fine-structure constant of the electromagnetic interaction $\alpha \approx 1/137$ and the electron mass $m^{}_e \approx 0.511~{\rm MeV}$. At this point, we do not need to know the inner structure of the proton, and the existence of the top quark and the weak gauge bosons. That is to say, the energy levels of a hydrogen atom can be calculated by neglecting all dynamics above the energy or momentum scale $\Lambda$ much higher than $\alpha m^{}_e$, and the corresponding error in the calculation can be estimated as $\alpha m^{}_e/\Lambda$. If a higher accuracy is required, $\Lambda$ will increase and the dynamics at a higher energy scale may be needed. See refs~\onlinecite{Manohar:1996cq,Pich:1998xt} for general reviews on effective field theories.

Now, consider a toy model with a light particle $\phi$ and a heavy one $\Phi$, whose masses are denoted by $m$ and $M$, respectively. Since $m \ll M$, there exist two widely separated energy scales. The Lagrangian for the full theory can be written as ${\cal L}^{}_{\rm full} = {\cal L}^{}_{\rm l}(\phi) + {\cal L}^{}_{\rm h}(\phi, \Phi)$, where the interaction between the light and heavy particles has been included in the second term. Since we are interested in physical phenomena at a low-energy scale $\mu \sim m \ll M$, where the experiments are carried out, we can integrate out the heavy particle as in the path-integral formalism. Hence, an effective Lagrangian involving only the light particle ${\cal L}^{}_{\rm eff} = {\cal L}^{}_{\rm l}(\phi) + \delta {\cal L}(\phi)$ is derived, and higher-dimensional operators appear in $\delta {\cal L}(\phi) = c^{}_i {\cal O}^{}_i/M^{d^{}_i-4}$, where $c_i$ is a coefficient and $d^{}_i$ stands for the mass dimension of ${\cal O}^{}_i$. It is evident that the dynamics at the high-energy scale can affect the low-energy physics by modifying the coupling constants and imposing symmetry constraints, but the overall effects are suppressed by the heavy particle mass $M$. The method of effective field theories becomes indispensable when we even do not know at all whether a complete theory with the heavy particles exists or not.

\vspace{5mm}

\noindent {\bf Matching and threshold effects.} Ultraviolet divergences appear in quantum field theories if radiative corrections are taken into account. In the presence of higher-dimensional operators, effective theories are nonrenormalizable in the sense that an ultraviolet divergence cannot be removed by a finite number of counter terms in the original Lagrangian. However, since there is an infinite number of higher-dimensional operators in the Lagrangian ${\cal L}^{}_{\rm eff}$, it is always possible to absorb all divergences and obtain finite results with a desired accuracy. Although any physical observables should be independent of the renormalization scheme used, it is a nontrivial task to choose a convenient renormalization scheme such that perturbative calculations are valid and simple.

According to the Appelquist--Carazzone theorem~\cite{Appelquist:1974tg}, heavy particles decouple automatically in a mass-dependent scheme, and their impact on the effective theory will be inversely proportional to the heavy particle mass $M$ and disappear in the limit of an infinitely large mass. Nevertheless, higher-order calculations in this scheme become quite involved. The mass-independent schemes, such as the modified minimal subtraction scheme ($\overline{\rm MS}$)~\cite{'tHooft:1973mm,'tHooft:1973us,Weinberg:1951ss}, have been suggested for practical computations in effective theories~\cite{Weinberg:1980wa}, where the strategy to construct a self-consistent effective theory in the $\overline{\rm MS}$ scheme is outlined and applied to the determination of the heavy gauge boson mass $M^{}_{\rm G}$ in the SU(5) grand unified theory~\cite{Hall:1980kf}.

One problem for the mass-independent scheme is that the heavy particles contribute equally to the so-called beta functions for gauge coupling constants, leading to an incorrect evolution at a low-energy scale $\mu \ll M^{}_{\rm G}$. The solution to this problem is to decouple heavy particles by hand and match the effective theory with the full theory at $\mu = M^{}_{\rm G}$ so that the same physical results can be produced in the effective theory as in the full theory. At any other energy scale below $M^{}_{\rm G}$, gauge coupling constants are governed by renormalization group running, which will be discussed in the following subsection. Therefore, if there are several heavy particles with very different masses, we should decouple them one by one to obtain a series of effective theories. The matching conditions (that is, the boundary conditions) at each mass scale are crucial for the effective theory to work below this scale. As a consequence, physical quantities (such as coupling constants and masses) may dramatically change at a decoupling scale or mass threshold. To figure out threshold effects, one has first to start with the full theory and construct the effective theories following the above strategy.

\newpage 

\noindent {\bf Renormalization group running.} The renormalization group was invented in 1953 by St\"{u}ckelberg and Petermann~\cite{Stueckelberg:1953dz}. However, it was Gell-Mann and Low~\cite{GellMann:1954fq} who studied the short-distance behaviour of the photon propagator in quantum electrodynamics in 1954 by using the renormalization group approach. The important role played by the renormalization group in Gell-Mann and Low's work was clarified in 1956 by Bogoliubov and Shirkov~\cite{Bogolyubov:1956gh}. The same approach was applied by Wilson to study critical phenomena and explain how phase transitions take place~\cite{Wilson:1971bg,Wilson:1971dh,Wilson:1973jj}.

The essential idea of the renormalization group stems from the fact that the theory is invariant under the change of renormalization prescription. More explicitly, if the theory is renormalized at a mass scale $\mu$, any change of $\mu$ will be compensated by changes in the renormalized coupling constant $g(\mu)$ and the mass $m(\mu)$ such that the theory remains the same. By requiring that the physical quantities, for example, the $S$-matrix element $S[\mu, g(\mu), m(\mu)]$, are invariant under this transformation, namely $\mu {\rm d}S/{\rm d}\mu = 0$, one can derive
\begin{equation}
\mu \frac{\partial S}{\partial \mu} + \beta \frac{\partial S}{\partial g} - \gamma^{}_{m} m \frac{\partial S}{\partial m} = 0 \; ,
\label{eq:CS}
\end{equation}
which is a specific form of the Callan--Symanzik equation~\cite{Callan:1970yg,Symanzik:1970rt}. Note that we have introduced the renormalization-group equations (RGEs) for the coupling constant and the mass
\begin{equation}
\mu \frac{\partial g(\mu)}{\partial \mu} = \beta(g) \; , ~~~~~~~~ - \frac{\mu}{m} \frac{\partial m(\mu)}{\partial \mu} = \gamma^{}_{m}(g) \; ,
\label{eq:running}
\end{equation}
where $\beta(g)$ and $\gamma^{}_{m}(g)$ are the beta function and the anomalous dimension, respectively, depending only on the coupling constant $g$ in the $\overline{\rm MS}$ scheme.

As pointed out by Weinberg a long time ago~\cite{Weinberg:1979sa}, the standard electroweak model can be regarded as an effective theory at low energies, and the impact of new physics at high-energy scales can be described by higher-dimensional operators, which are composed of the already known SM fields. If the SM gauge symmetry is preserved, but the accidental symmetry of lepton number is violated, there will be a unique dimension-five operator ${\cal O}^{}_5 = \overline{\ell^{}_{\rm L}} H H^{\rm T} \ell^C_{\rm L}$, where $\ell^{}_{\rm L}$ and $H$ stand for the SM lepton and Higgs doublets, respectively. After spontaneous breakdown of electroweak gauge symmetry, neutrinos acquire finite masses from the so-called Weinberg operator ${\cal O}^{}_5$. Therefore, neutrinos are assumed to be Majorana particles in this case. It is expected that the lightness of neutrinos can be ascribed to the existence of a superhigh-energy scale. Now, it becomes clear that if neutrino masses originate from some dynamics at a high-energy scale, such as the GUT scale, neutrino parameters including leptonic mixing parameters and neutrino masses will evolve according to their RGEs as the energy scale goes down to where the parameters are actually measured in low-energy experiments.

\section*{Neutrino mass models}
\label{sec:neu_mass_mod}

\begin{figure}[t]
\includegraphics[width=0.95\textwidth]{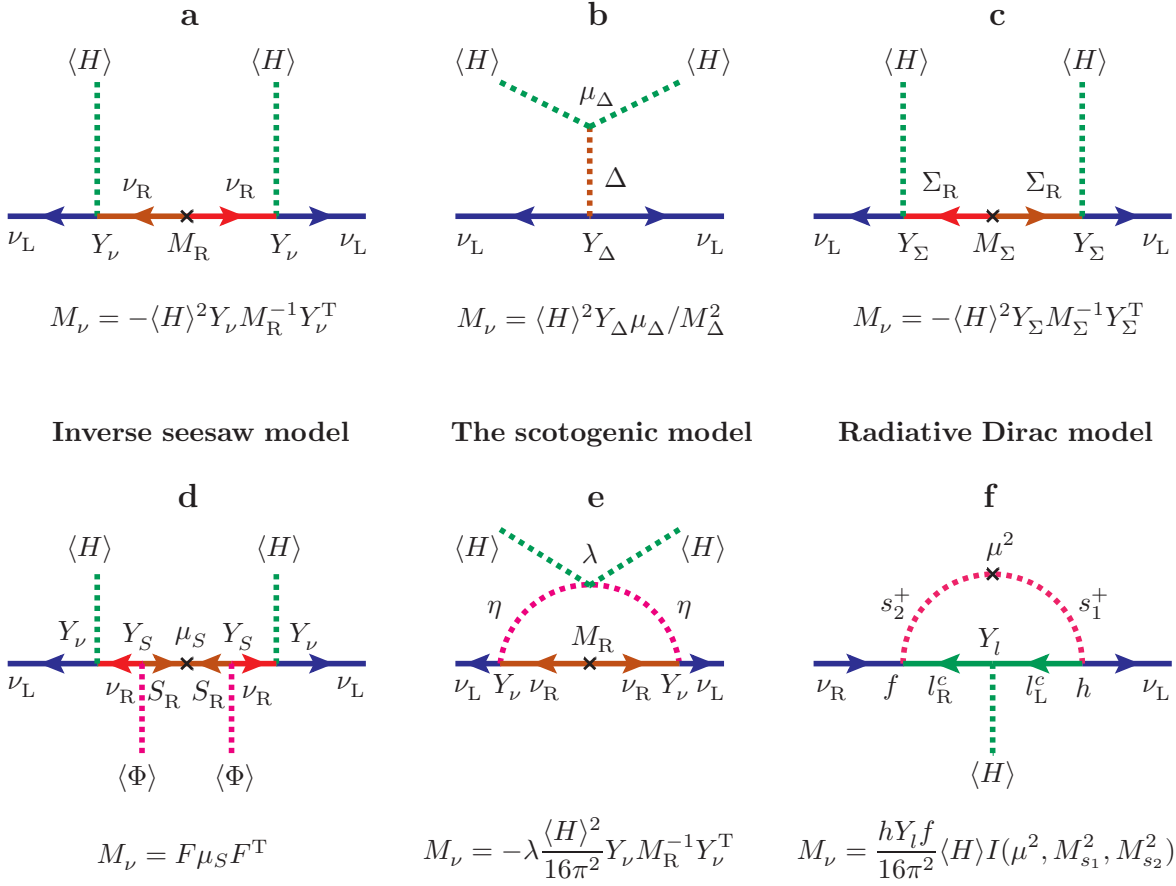}
\vspace{0.5cm} \caption{{\bf Feynman diagrams for neutrino mass generation.} The origin of neutrino masses calls for new physics beyond the Standard Model of particle physics. Each neutrino mass model is represented by a Feynman diagram, below which the effective neutrino mass matrix $M^{}_\nu$ is given. In the canonical seesaw models of type-I {\bf (a)}, type-II {\bf (b)}, and type-III {\bf (c)}, after the new super-heavy particles are integrated out, the dimension-five Weinberg operator is obtained and neutrinos acquire tiny Majorana masses. {\bf (d)}~In the inverse seesaw model, the lepton-number-violating $\mu^{}_S$ term can be naturally at the ${\rm keV}$ scale, which together with $F \equiv Y^{}_\nu \langle H \rangle [Y^{\rm T}_S \langle \Phi \rangle]^{-1} \sim {\cal O}(10^{-2})$ leads to sub-eV neutrino masses. {\bf (e)}~In the scotogenic model, neutrino masses are suppressed by a loop factor compared to the canonical type-I seesaw model. Hence, the masses of right-handed neutrinos and another Higgs doublet can be at the TeV scale, which is accessible at the Large Hadron Collider at CERN, and additional scalar bosons could be dark-matter candidates. {\bf (f)}~This is an example to show how to realize Dirac neutrino masses~\cite{Kanemura:2011jj}. Two singly-charged scalar particles $s^+_{1,2}$ with masses $M^{}_{s_{1,2}}$ are added, and they are mixed through a mass term $\mu^2$. Compared to the charged-lepton masses, neutrino masses are suppressed by the loop factor, Yukawa couplings $h$ and $f$, and perhaps the loop function $I(\mu^2, M^2_{s_1}, M^2_{s_2})$ that depends on the mass parameters.}
\label{fig:diagrams}
\end{figure}

\noindent To generate tiny neutrino masses, one has to go beyond the SM and extend its particle content, or its symmetry structure, or both. In this section, we summarize several typical neutrino mass models, which are natural extensions of the SM that have attracted a lot of attention in the past decades. In Fig.~\ref{fig:diagrams}, the Feynman diagrams for neutrino mass generation in those models are shown.

\label{sub:seesaw}
\vspace{5mm}

\noindent {\bf Canonical seesaw models.} As the Higgs particle has recently been discovered in the ATLAS~\cite{Aad:2012tfa} and CMS~\cite{Chatrchyan:2012ufa}
experiments at the Large Hadron Collider, the SM gauge
symmetry $SU(2)_{\rm L} \times U(1)_{\rm Y}$ and its
spontaneous breaking via the Higgs mechanism seem to work perfectly
in describing the electromagnetic and weak interactions. On the
other hand, nonzero neutrino masses indicate that the SM may just be
an effective theory below and around the electroweak scale
$\Lambda^{}_{\rm EW} = 10^2~{\rm GeV}$. Thus, one can preserve the
SM gauge symmetry structure and take into account all
higher-dimensional operators, which are relevant for neutrino
masses, as pointed out by Weinberg \cite{Weinberg:1979sa}. The total Lagrangian is
\begin{equation}
{\cal L} = {\cal L}^{}_{\rm SM} - \left[ \frac{1}{2} \kappa^{}_{\alpha \beta} (\overline{\ell^{}_{\alpha {\rm L}}} H)
(H^{\rm T} \ell^C_{\beta {\rm L}}) + {\rm h.c.} \right] \; ,
\label{eq:SS1}
\end{equation}
where ${\cal L}^{}_{\rm SM}$ denotes the SM Lagrangian, $\ell^{}_{\alpha {\rm L}}$ and $H$ stand for the SM lepton and Higgs doublets, respectively. The coefficients $\kappa^{}_{\alpha
\beta}$ ($\alpha, \beta = e, \mu, \tau$) are of mass-dimension
$-1$ and related to the Majorana neutrino mass matrix as $M^{}_\nu =
\kappa \langle H \rangle^2$, where $\langle H \rangle \approx
174~{\rm GeV}$ is the vacuum expectation value of $H$.

One of the simplest extensions of the SM, leading to the Weinberg
operator, is the so-called type-I seesaw model, in which three right-handed singlet neutrinos $\nu^{}_{\rm R}$ are introduced. Since the $\nu^{}_{\rm R}$'s are neutral under transformations of the SM gauge symmetry, they can have Majorana mass terms, namely, their masses are the eigenvalues of a complex and symmetric mass matrix $M^{}_{\rm R}$. On the other hand, they are coupled to the lepton and Higgs doublets via a Yukawa-type interaction with a coupling matrix $Y^{}_\nu$. Since the masses of right-handed neutrinos are not subject to electroweak symmetry breaking, we can assume that ${\cal O}(M^{}_{\rm R}) \gg \Lambda^{}_{\rm EW}$ and integrate out the three $\nu^{}_{\rm R}$'s.
At a lower-energy scale, one obtains the Weinberg operator with $\kappa = - Y^{}_\nu M^{-1}_{\rm R} Y^{\rm T}_\nu$. Therefore, the smallness of neutrino masses can be attributed to the heaviness of the $\nu^{}_{\rm R}$'s~\cite{Minkowski:1977sc,Yanagida:1979ss,Gell-Mann:1979ss,Glashow:1979ss,Mohapatra:1979ia}.

In the type-II seesaw model~\cite{type2,Valle,Cheng,Magg,Shafi,Mohapatra}, the scalar sector of the SM is enlarged with a Higgs triplet $\Delta$. To avoid an unwanted Goldstone boson associated with the spontaneous breakdown of the global $U(1)$ lepton number symmetry, one can couple the Higgs triplet to the lepton doublet with a Yukawa coupling matrix $Y^{}_\Delta$, and simultaneously to the Higgs doublet with a mass parameter $\mu^{}_\Delta$. Assuming that the Higgs triplet mass $M^{}_\Delta$ is well above the electroweak scale, that is, $M^{}_\Delta \gg \Lambda^{}_{\rm EW}$, we can integrate out $\Delta$ to obtain the Weinberg operator with $\kappa = Y^{}_\Delta \mu^{}_\Delta/M^2_\Delta$, indicating that the neutrino masses are suppressed by $M^{}_\Delta$.

In the type-III seesaw model~\cite{type3}, one introduces three fermion triplets $\Sigma^{}_i$ ($i = 1, 2, 3$) and couple them to the lepton and Higgs doublets with a Yukawa coupling matrix $Y^{}_\Sigma$. In each $\Sigma^{}_i$, there are three heavy fermions: two charged fermions $\Sigma^\pm_i$ and one neutral fermion $\Sigma^0_i$. Given a Majorana mass matrix $M^{}_\Sigma$ of the fermion triplets and ${\cal O}(M^{}_\Sigma) \gg \Lambda^{}_{\rm EW}$, we can construct an effective theory without the heavy $\Sigma^{}_i$'s at a lower-energy scale. In this effective theory, the same Weinberg operator for neutrino masses can be obtained and the coefficient is identified as $\kappa = - Y^{}_\Sigma M^{-1}_{\Sigma}Y^{\rm T}_\Sigma$. One can observe that the $\Sigma^0_i$'s are playing the same role in generating neutrino masses as the $\nu^{}_{\rm R}$'s in the type-I seesaw model. However, due to their gauge interaction, the fermion triplets are subject to more restrictive constraints from lepton-flavour-violating decays of charged leptons and direct collider searches.

A common feature of the above three seesaw models is the existence of superheavy particles. Given neutrino masses ${\cal O}(M^{}_\nu) \sim 0.1~{\rm eV}$ and $\langle H \rangle \sim 100~{\rm GeV}$, one can estimate the seesaw scale $\Lambda^{}_{\rm SS} \sim 10^{14}~{\rm GeV}$. Therefore, an effective theory with the same Weinberg operator is justified at any scale between $\Lambda^{}_{\rm EW}$ and $\Lambda^{}_{\rm SS}$. Although the leptogenesis mechanism for the matter--antimatter asymmetry can be perfectly implemented in the seesaw framework, the heaviness of new particles renders the seesaw models difficult to be tested in low-energy and collider experiments.

\label{sub:inverse}
\vspace{5mm}

\noindent {\bf Inverse seesaw model.} To lower the typical seesaw scale $\Lambda^{}_{\rm SS}$ in a natural way, one can extend the type-I seesaw model by adding three right-handed singlet fermions $S^{}_{\rm R}$ and one Higgs singlet $\Phi$, both of which are coupled to the $\nu^{}_{\rm R}$'s by a Yukawa coupling matrix $Y^{}_S$. A proper assignment of quantum numbers under a specific global symmetry can be used to forbid the Majorana mass term of $\nu^{}_{\rm R}$ and the $\nu^{}_{\rm R}$-$\Phi$ Yukawa interaction. However, the mixing between $\nu^{}_{\rm R}$ and $S^{}_{\rm R}$ is allowed through a Dirac mass term $M^{}_S = Y^{}_S \langle \Phi \rangle$, so is the Majorana mass term $\mu^{}_S \overline{S^C_{\rm R}} S^{}_{\rm R}$. In this setup, the Majorana mass matrix for three light neutrinos is given by $M^{}_\nu = M^{}_{\rm D} (M^{\rm T}_S)^{-1} \mu^{}_S M^{-1}_S M^{\rm T}_{\rm D}$, where $M^{}_{\rm D} \equiv Y^{}_\nu \langle H \rangle$ is the Dirac neutrino mass matrix as in the type-I seesaw model.

Given ${\cal O}(M^{}_S) \sim 10~{\rm TeV}$ and ${\cal O}(M^{}_{\rm D}) \sim 10^2~{\rm GeV}$, the sub-eV neutrino masses ${\cal O}(M^{}_\nu) \sim 0.1~{\rm eV}$ can be achieved by assuming $\mu^{}_S \sim 1~{\rm keV}$. In this inverse seesaw model~\cite{inverse}, the neutrino masses are not only suppressed by the ratio of the electroweak and seesaw energy scales, that is, $\Lambda^{}_{\rm EW}/\Lambda^{}_{\rm SS} = M^{}_{\rm D}/M^{}_S \sim 10^{-2}$, but also by the tiny lepton-number-violating mass parameter $\mu^{}_S$ compared to the ordinary seesaw scale. The smallness of $\mu^{}_S$ is natural in the sense that the model preserves the lepton number symmetry in the limit $\mu^{}_S \to 0$ (ref.~\onlinecite{tHooft}). In contrast to the ordinary seesaw models, the inverse seesaw model is testable through non-unitarity effects in neutrino oscillation experiments~\cite{Malinsky:2009df}, lepton-flavour-violating decays of charged leptons~\cite{Deppisch:2004fa,Deppisch:2005zm,Ibanez:2009du}, and collider experiments~\cite{Hirsch:2009ra,Das:2012ze}.

\vspace{5mm}

\noindent {\bf Scotogenic model.} A radiative mechanism for neutrino mass generation is to attribute the smallness of neutrino masses to loop suppression instead of the existence of superheavy particles~\cite{Zee:1980ai,Zee:1985id,Babu:1988ki,Ma:2006km,Gustafsson:2012vj}. One interesting model of this type is the so-called scotogenic model~\cite{Ma:2006km}, where three $\nu^{}_{\rm R}$'s and one extra Higgs doublet $\eta$ are added to the SM. Furthermore, a $Z^{}_2$ symmetry is imposed on the model such that all SM fields are even, while $\nu^{}_{\rm R}$ and $\eta$ are odd. Even though the $SU(2)^{}_{\rm L} \times U(1)_{\rm Y}$ quantum numbers of $\eta$ are the same as the SM Higgs doublet and the $\nu^{}_{\rm R}$'s have a Majorana mass term, the Dirac neutrino mass term is forbidden by the $Z^{}_2$ symmetry and neutrino masses are vanishing at tree level.

In the scotogenic model, neutrino masses appear first at one-loop level and the exact $Z^{}_2$ symmetry guarantees the stability of one neutral scalar boson (from the Higgs doublet $\eta$), which would be a good candidate for a dark matter particle~\cite{Ma:2006km}. Due to loop suppression, sub-eV neutrino masses can be obtained even when $\nu^{}_{\rm R}$'s and scalar particles are at the TeV scale. Therefore, this model has observable effects in lepton-flavour-violating processes, relic density of dark matter, and collider phenomenology~\cite{Ma:2006km}.

\vspace{5mm}

\noindent {\bf Dirac neutrino model.} Finally, we consider the Dirac neutrino model. In the SM model, both quarks and charged leptons acquire their masses through Yukawa interactions with the Higgs doublet. After introducing three $\nu^{}_{\rm R}$'s, one can do exactly the same thing for neutrinos, and thus, tiny neutrino masses can be ascribed to the smallness of neutrino Yukawa couplings. One difficulty with the Dirac neutrino model is why the fermion masses span twelve orders of magnitude, exaggerating the strong hierarchy problem of fermion masses in the SM. Solutions to the above problem can be found in extra-dimensional models~\cite{Dienes:1998sb}, where the SM particles are confined to a three-dimensional brane and the $\nu^{}_{\rm R}$'s are allowed to feel one or more extra dimensions~\cite{ArkaniHamed:1998vp}. In this case, the neutrino Yukawa couplings are highly suppressed by the large volume of the extra dimensions. Another solution is to implement a radiative mechanism, as in the scotogenic model, such that light neutrino masses are due to loop suppression~\cite{Branco:1978bz,Chang:1986bp,Hung:1998tv,Kanemura:2011jj}. See Fig.~\ref{fig:diagrams} for an illustration. However, in both kinds of models, an additional $U(1)$ symmetry (that is, lepton number conservation) has to be enforced to forbid a Majorana mass term.

\section*{Running behaviour of neutrino parameters}
\label{sec:running}

\noindent Now, we proceed to discuss the running behaviour of neutrino parameters. First of all, we note that there are two different ways to study the renormalization group running. In the top--down scenario, a full theory is known at the high-energy scale and the theoretical predictions for neutrino parameters are given as initial conditions. At the threshold of heavy particle decoupling, one has to match the resulting effective theory with the full theory, so that the unknown parameters in the effective theory can be determined and used to reproduce the same physical results as in the full theory. Then, the running is continued in the effective theory. This procedure should be repeated in the case of multiple particle thresholds until a low-energy scale where the neutrino parameters are measured. In the bottom--up scenario, we start with the experimental values of neutrino parameters at a low-energy scale, and evolve them by using the RGEs in the effective theory to the first particle threshold. At this moment, more input or assumptions about the dynamics above the threshold are needed for the running to continue. Otherwise, the running is terminated and some useful information on the full theory cannot be obtained.

In the following, we will focus on the bottom--up approach and explore the implications of measurements of neutrino parameters for the dynamics at a high-energy scale, where a full theory of neutrino masses and lepton flavour mixing may exist. However, we shall also comment on the threshold effects in the top--down scenario once a specific flavour model is assumed. In the effective theory, where the SM is extended by the Weinberg operator, the RGE for the effective neutrino mass parameter $\kappa$ was first derived in refs~\onlinecite{Chankowski:1993tx,Babu:1993qv}, and revised in ref.~\onlinecite{Antusch:2001ck}. In general, we have the RGE for $\kappa$ given by
\begin{equation}
16\pi^2 \frac{{\rm d} \kappa}{{\rm d}t} = \alpha^{}_\kappa \kappa + C^{}_\kappa \left[\left(Y^{}_l Y^\dagger_l\right)\kappa + \kappa \left(Y^{}_l Y^\dagger_l\right)^{\rm T}\right] \; ,
\label{eq:RGEkappa}
\end{equation}
where $t = \ln(\mu/\Lambda_{\rm EW})$ and $Y^{}_l$ stands for the charged-lepton Yukawa coupling matrix. In equation~(\ref{eq:RGEkappa}), $C^{}_\kappa$ is a constant, while $\alpha^{}_\kappa$ depends on the gauge couplings and all Yukawa coupling matrices of the charged fermions. Given initial values of all relevant coupling constants and masses at $\Lambda^{}_{\rm EW}$, one can evaluate the neutrino parameters at any energy scale between $\Lambda^{}_{\rm EW}$ and a cutoff scale $\Lambda$, after solving equation~(\ref{eq:RGEkappa}) together with the RGEs of the other model parameters and diagonalizing $\kappa$. Since $\kappa$ is diagonalized by the lepton flavour mixing matrix $U(\theta^{}_{12}, \theta^{}_{13}, \theta^{}_{23}, \delta, \rho, \sigma)$ in the basis where $Y^{}_l$ is diagonal, one can derive, using equation~(\ref{eq:RGEkappa}), the individual RGEs for the leptonic mixing angles $\{\theta^{}_{12}, \theta^{}_{13}, \theta^{}_{23}\}$, the CP-violating phases $\{\delta, \rho, \sigma\}$, and the neutrino mass eigenvalues $\{m^{}_1, m^{}_2, m^{}_3\}$, which can be found in refs~\onlinecite{Antusch:2003kp,Chankowski:2001mx,Ray:2010rz}.

\vspace{5mm}

\noindent {\bf The Standard Model.} In the framework of the SM, the relevant coefficients in equation~(\ref{eq:RGEkappa}) are given by $C^{\rm SM}_\kappa = -3/2$ and $\alpha^{\rm SM}_\kappa \approx -3g^2_2 + 2y^2_\tau  + 6 y^2_t + \lambda$, where only the Yukawa couplings of the heaviest charged lepton and quark are retained, and $\lambda$ is the quartic Higgs self-coupling constant. Since the Yukawa couplings of charged leptons are small compared to gauge couplings, the evolution of neutrino masses can be essentially described by a common scaling factor. For the running of the leptonic mixing angles, the contribution from tau Yukawa coupling $y^{}_\tau = m^{}_\tau/\langle H \rangle \sim 0.01$ is dominant. However, $y^{}_\tau$ itself is already a very small number, so one expects that the running effects of all three leptonic mixing angles are generally insignificant.

\begin{figure}[t]
\includegraphics[width=0.8\textwidth]{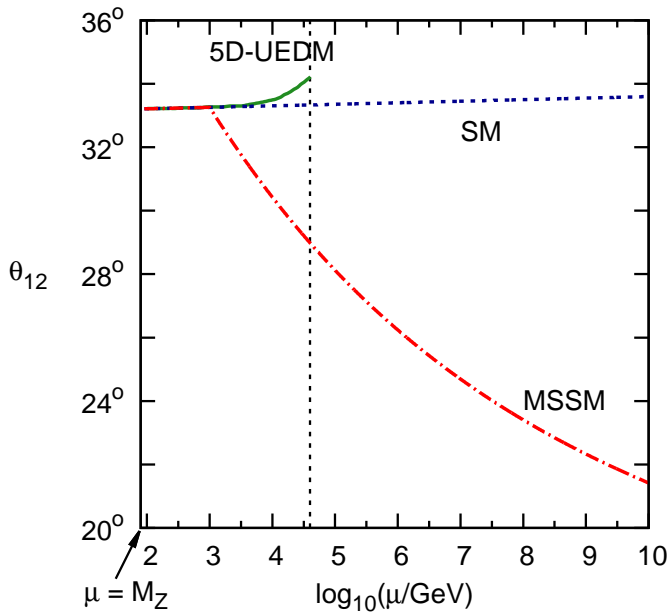}
\vspace{-1.5mm} \caption{{\bf Evolution of the leptonic mixing angle $\theta^{}_{12}$ in the bottom--up scenario.} The running behaviour of $\theta^{}_{12}$ is compared to the Standard Model (SM, dotted curve), the Minimal Supersymmetric Standard Model (MSSM, dotted-dashed curve), and the five-dimensional universal extra-dimensional model (5D-UEDM, solid curve), where the vertical dashed line corresponds to a cutoff scale $\Lambda = 40~{\rm TeV}$. In this figure, a quasi-degenerate neutrino mass spectrum is assumed and $m^{}_1 = 0.2~{\rm eV}$ is chosen. At the energy scale $M^{}_Z = 91.19~{\rm GeV}$, the gauge coupling constants and the quark mixing parameters are taken from the Particle Data Group~\cite{PDG:2012}, the quark and charged-lepton masses from refs~\onlinecite{Xing:2007fb,Xing:2011aa}, and the leptonic mixing parameters are set to the best-fit values of the NuFit group~\cite{GonzalezGarcia:2012sz}. In addition, a supersymmetry breaking scale at $1~{\rm TeV}$ and $\tan \beta = 10$ are assumed in the MSSM case.  \label{fig:theta12}}
\end{figure}

On the other hand, the evolution of the leptonic mixing angles can be enhanced if the neutrino mass spectrum is quasi-degenerate, that is, $m^2_i \gg |\Delta m^2_{31}|$. In particular, the leptonic mixing angle $\theta^{}_{12}$ has the strongest running effects, partly due to $\Delta m^2_{21} \ll |\Delta m^2_{31}|$. In the limit of quasi-degenerate mass spectrum and CP conservation, the RGEs for the two neutrino mass-squared differences and the three leptonic mixing angles are given by
\begin{align}
8\pi^2 \displaystyle \frac{\rm d}{{\rm d}t} \Delta m^2_{21} &\approx \alpha^{}_\kappa \Delta m^2_{21} + C^{}_\kappa y^2_\tau \left[2s^2_{23}(m^2_2 c^2_{12}- m^2_1 s^2_{12}) + (m^2_1 + m^2_2)\sin 2\theta^{}_{23} \sin 2\theta^{}_{12} s^{}_{13} \right] \; , \\
8\pi^2 \displaystyle \frac{\rm d}{{\rm d}t} \Delta m^2_{31} &\approx \alpha^{}_\kappa \Delta m^2_{31} + C^{}_\kappa y^2_\tau \left[2(m^2_3 c^2_{23} - m^2_1 s^2_{12} s^2_{23}) + m^2_1 \sin 2\theta^{}_{23} \sin 2\theta^{}_{12} s^{}_{13} \right] \; , \\
8\pi^2 \displaystyle \frac{\rm d}{{\rm d}t} \theta^{}_{12} &\approx - C^{}_\kappa y^2_\tau \displaystyle \frac{m^2_1}{\Delta m^2_{21}} s^2_{23} \sin 2\theta^{}_{12} \; , \\
8\pi^2 \displaystyle \frac{\rm d}{{\rm d}t} \theta^{}_{13} &\approx - C^{}_\kappa y^2_\tau \displaystyle \frac{m^2_1}{\Delta m^2_{31}} c^2_{23} \sin 2\theta^{}_{13} \; , \\
8\pi^2 \displaystyle \frac{\rm d}{{\rm d}t} \theta^{}_{23} &\approx - C^{}_\kappa y^2_\tau \displaystyle \frac{m^2_1}{\Delta m^2_{31}} \sin 2\theta^{}_{23} \; ,
\end{align}
where the relevant coefficients are presented in Table~\ref{tab:RGE_coeff}.
\begin{table}[t]
\caption{{\bf Relevant coefficients of the renormalization group equations for neutrino parameters.}}
\label{tab:RGE_coeff}
\renewcommand{\arraystretch}{2.0}
\label{tab:coefficients}
\begin{center}
\begin{tabular}{lccc}
\quad \quad ~ \quad \quad & \quad \quad  \quad
SM\footnote{Here $g^{}_2$ and $g^{}_1$ are the $SU(2)_{\rm L} \times U(1)_{\rm Y}$ gauge couplings, $\lambda$ the Higgs self-coupling constant, $y^{}_\tau$ and $y^{}_t$ the Yukawa couplings of tau charged lepton and top quark, respectively.} \quad \quad   \quad & \quad \quad   \quad
MSSM \quad \quad \quad &  \quad \quad  \quad
5D-UEDM\footnote{At the energy scale $\mu$, the number of excited Kaluza--Klein states $s = \lfloor\mu/\mu^{}_0\rfloor$ is an integer just below $\mu/\mu^{}_0$ with $\mu^{}_0 = 1~{\rm TeV}$. } \quad \quad \quad
\\
\hline
\multirow{1}{*}{$C^{}_\kappa$}   & $-3/2$ & $1$ & $-3(1+s)/2$
\vspace{0.2cm}
\\
\multirow{2}{0.8cm}{$\alpha^{}_\kappa$}   & $-3g^2_2 + 2y^2_\tau  $ & $-6 g^2_1 / 5 - 6 g^2_2 $ &
$-3g^2_2 + 2y^2_\tau  + 6 y^2_t + \lambda $
\\ & $+ 6 y^2_t + \lambda$ & $+ 6 y^2_t$ & $ ~+ s (- 3 g^2_1/20 - 11 g^2_2/4 + 12 y^2_t + \lambda)$~
\vspace{0.2cm}
\\
\hline 
\end{tabular}
\end{center}
\end{table}
It is now straightforward to observe that the evolution of $\theta^{}_{12}$ is enhanced by a factor of $|\Delta m^2_{31}|/\Delta m^2_{21} \approx 30$, compared to that of $\theta^{}_{13}$ and $\theta^{}_{23}$. For illustration, the evolution of $\theta^{}_{12}$ from $M^{}_Z = 91.19~{\rm GeV}$ to $\Lambda = 10^{10}~{\rm GeV}$ is shown in Fig.~\ref{fig:theta12}. At $M^{}_Z$, the gauge coupling constants and the quark mixing parameters are taken from the Particle Data Group~\cite{PDG:2012}, the quark and charged-lepton masses from refs~\onlinecite{Xing:2007fb,Xing:2011aa}, and the leptonic mixing parameters are set to the best-fit values from the NuFit group~\cite{GonzalezGarcia:2012sz}. The Higgs mass $M^{}_H = 126~{\rm GeV}$ is assumed to be consistent with the latest measurements by the ATLAS~\cite{Aad:2012tfa} and CMS~\cite{Chatrchyan:2012ufa} experiments.
It is worthwhile to mention that $M_H$ or equivalently the Higgs self-coupling constant $\lambda = M^2_H/\langle H \rangle^2$ affects the running of neutrino masses, and also the SM vacuum stability~\cite{Buttazzo:2013uya}. Finally, a quasi-degenerate neutrino mass spectrum is adopted with the lightest neutrino mass $m^{}_1 = 0.2~{\rm eV}$ and the Majorana CP-violating phases $\{\rho, \sigma\}$ are set to zero. Even with these extremely optimistic assumptions, the value of $\theta^{}_{12}$ turns out to be only larger by $1^\circ$ at $\Lambda = 10^{10}~{\rm GeV}$ than at $M^{}_Z$.

The previous observations apply well to seesaw models with $\Lambda = 10^{10}~{\rm GeV}$ identified as the mass of the lightest new particle. Above the seesaw threshold, the running of neutrino parameters has also been studied in the complete type-I~\cite{King:2000hk,Antusch:2005gp,Mei:2005qp}, type-II~\cite{Chao:2006ye,Schmidt:2007nq,Joaquim:2009vp}, and type-III~\cite{Chakrabortty:2008zh} seesaw models. However, for low-scale neutrino mass models, there exist new particles at the TeV scale. Therefore, the running behaviour of neutrino parameters can be significantly changed by threshold effects in the inverse seesaw model~\cite{Bergstrom:2010qb,Bergstrom:2010id} and the scotogenic model~\cite{Bouchand:2012dx}. In the Dirac neutrino model, the RGEs of the neutrino parameters have also been derived and investigated in detail~\cite{Lindner:2005as}.

\vspace{5mm}

\noindent {\bf Supersymmetric models.} In the minimal supersymmetric extension of the SM (MSSM), all fermions have bosonic partners, and vice versa~\cite{Nilles:1983ge}. Although there is so far no direct hint on supersymmetry, the MSSM is regarded as one of the most natural alternatives to the SM for its three salient features: (1) elimination of the fine-tuning or hierarchy problem; (2) implication for grand unification of gauge coupling constants; (3) candidates for the dark matter. Hence, neutrino mass models in the supersymmetric framework are extensively studied in the literature~\cite{Antusch:2003kp}.

In the MSSM extended with the Weinberg operator, the corresponding coefficients in equation~(\ref{eq:RGEkappa}) are $C^{\rm MSSM}_\kappa = 1$ and $\alpha^{\rm MSSM}_\kappa \approx -6 g^2_1 / 5 - 6 g^2_2 + 6 y^2_t$. The neutrino mass matrix is then given by $M^{}_\nu = \kappa \langle H \rangle^2 \sin^2 \beta$
with $\tan \beta$ being the ratio of the vacuum expectation values of the two Higgs doublets in the MSSM. Similar to the SM, the running of the leptonic mixing angles is dominated by the tau Yukawa coupling $y^{}_\tau = m^{}_\tau \sqrt{1+\tan^2 \beta}/\langle H \rangle$. However, now $y^{}_\tau$ can be remarkably larger than its value in the SM if a large value of $\tan \beta$ is chosen. Consequently, apart from the enhancement due to a quasi-degenerate neutrino mass spectrum, the running effects of the leptonic mixing angles can be enlarged by $\tan \beta$. In Fig.~\ref{fig:theta12}, we show the evolution of $\theta^{}_{12}$ in the MSSM with $\tan \beta = 10$, where the input values at $M^{}_Z$ are the same as in the SM. In addition, the supersymmetry breaking scale is assumed to be $1~{\rm TeV}$, below which the SM works well as an effective theory. The value of $\theta^{}_{12}$ decreases with respect to an increasing energy scale, whereas it increases in the SM. This is due to the opposite signs of $C^{\rm MSSM}_\kappa$ and $C^{\rm SM}_\kappa$.

\begin{figure}[t]
\includegraphics[width=0.8\textwidth]{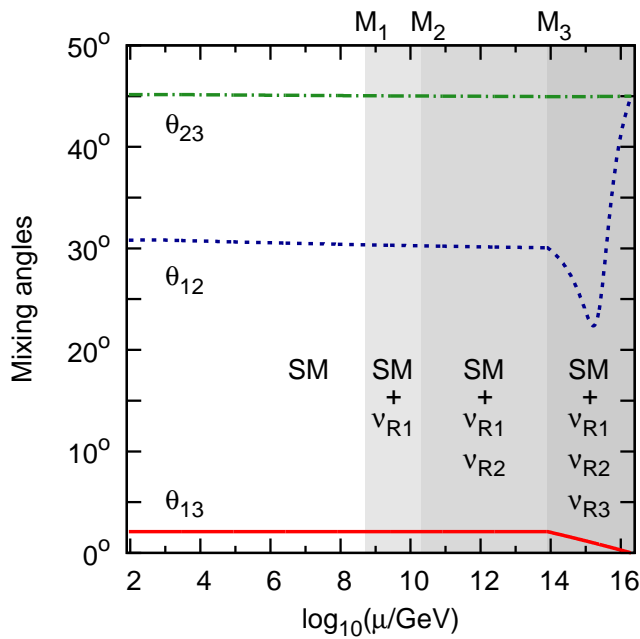}
\vspace{-1.5mm} \caption{{\bf Evolution of the leptonic mixing angles in the top--down scenario.} At the GUT scale, the bimaximal-mixing pattern with $\theta^{}_{12} = \theta^{}_{23} = 45^\circ$ and $\theta^{}_{13} = 0$ is assumed, and used to reconstruct the effective neutrino mass matrix $M^{}_\nu$. The mass matrix of heavy Majorana neutrinos $M^{}_{\rm R}$ is built from $M^{}_\nu$ and a specific structure of $Y^{}_\nu$ by inverting the seesaw formula~\cite{Antusch:2005gp}, namely $M^{}_{\rm R} = Y^{\rm T}_\nu M^{-1}_\nu Y^{}_\nu$. Consequently, the heavy neutrino masses $M^{}_3 = 8.1\times 10^{13}~{\rm GeV}$, $M^{}_2 = 2.1\times 10^{10}~{\rm GeV}$, and $M^{}_1 = 5.5\times 10^8~{\rm GeV}$ can be obtained, and they determine the decoupling energy scales represented by shaded regions, where the particle content in each energy region is also indicated. This figure is adapted, with permission, from ref.~\onlinecite{Antusch:2005gp} \copyright Institute of Physics. \label{fig:bimaximal}}
\end{figure}

As an example for the top--down approach, one considers a bimaximal-mixing pattern (that is, $\theta^{}_{12} = \theta^{}_{23} = 45^\circ$ and $\theta^{}_{13} = 0$) at the GUT scale $\Lambda^{}_{\rm GUT} = 2 \times 10^{16}~{\rm GeV}$ (refs~\onlinecite{Antusch:2005gp,Antusch:2002hy}). It is worthwhile to mention that the leptonic mixing angles above the seesaw scale arise from the diagonalization of $Y^{}_\nu M^{-1}_{\rm R} Y^{\rm T}_\nu$, and the leptonic mixing angles and the neutrino masses at this scale can be viewed as a convenient parametrization of $Y^{}_\nu M^{-1}_{\rm R} Y^{\rm T}_\nu$, which is a combination of fundamental model parameters $Y^{}_\nu$ and $M^{}_{\rm R}$. Therefore, a bimaximal-mixing pattern may result from a flavour symmetry at the GUT scale. For a complete type-I seesaw model at $\Lambda^{}_{\rm GUT}$, the full flavour structure of the neutrino Yukawa coupling matrix $Y^{}_\nu$ should be specified, and the mass matrix of right-handed neutrinos is reconstructed from the light neutrino mass matrix and $Y^{}_\nu$ from the seesaw formula. See ref.~\onlinecite{Antusch:2005gp} for the other input parameters. In Fig.~\ref{fig:bimaximal}, the running behaviour of the three leptonic mixing angles are depicted, where the gray-shaded areas stand for the decoupling of three right-handed neutrinos at $M^{}_3 = 8.1\times 10^{13}~{\rm GeV}$, $M^{}_2 = 2.1\times 10^{10}~{\rm GeV}$, and $M^{}_1 = 5.5\times 10^8~{\rm GeV}$. As one can observe from Fig.~\ref{fig:bimaximal}, the decoupling of the heaviest right-handed neutrino and the matching between the first effective theory and the full theory have remarkable impact on the running of $\theta^{}_{12}$ and $\theta^{}_{13}$. This impact depends on the presumed flavour structure in the lepton sector, indicating that the running of neutrino parameters has to be taken into account in the flavour model at a super-high energy scale.

In the MSSM, it is in general expected that running effects of neutrino parameters are significant, in particular for large values of $\tan \beta$ and a quasi-degenerate neutrino mass spectrum. This generic feature should also be applicable to supersymmetric versions of neutrino mass models discussed in the previous section.

\vspace{5mm}

\noindent {\bf Extra-dimensional models.} The existence of one or more extra spatial dimensions was first considered by Kaluza~\cite{Kaluza:1921tu} and Klein~\cite{Klein:1926tv} in the 1920s. The recent interest in extra dimensions and their implications for particle physics was revived by the seminal works in refs~\onlinecite{ADD,RS1,RS2}. In extra-dimensional models, the fundamental energy scale for gravity can be as low as a few TeV, solving the gauge hierarchy problem of the SM. Furthermore, the excited Kaluza--Klein (KK) modes of the SM fields serve as promising candidates for cold dark matter. See ref.~\onlinecite{Randall}, for a brief review.

As an interesting example for the running of neutrino parameters in extra-dimensional models, we consider the so-called universal extra-dimensional model (UEDM) first introduced in ref.~\onlinecite{Appelquist:2000nn}, in which all SM fields are allowed to propagate in one or more compact extra dimensions. Since the KK number is conserved and the excited KK modes manifest themselves only at loop level, current mass bound on the first KK excitation from electroweak precision measurements and direct collider searches is just about a few hundred GeV (ref.~\onlinecite{Appelquist:2000nn}). In the five-dimensional UEDM, the corresponding coefficients in equation~(\ref{eq:RGEkappa}) are $C^{\rm UEDM}_\kappa = (1+s) C^{\rm SM}_\kappa$ and $\alpha^{\rm UEDM}_\kappa = \alpha^{\rm SM}_\kappa + s (- 3 g^2_1/20 - 11 g^2_2/4 + \lambda + 12 y^2_t)$, where $s = \lfloor \mu/ \mu^{}_0 \rfloor$ is the number of excited KK modes at the energy scale $\mu$. Note that $\mu^{}_0$ denotes the mass of the first KK excitation, or equivalently $R = \mu^{-1}_0$ is the radius of the compact extra dimension. In contrast to the SM and the MSSM, the running of $\kappa$ in the UEDM obeys a power law due to the increasing number of excited KK modes, implying a significant boost in the running~\cite{Blennow:2011mp,Ohlsson:2012hi,Cornell:2012qf}. The reason is simply that, at a given energy scale $\mu$, we have an effective theory with $s = \lfloor \mu/ \mu^{}_0 \rfloor$ new particles, which will run in the loops and contribute to the RGEs of neutrino parameters.

In Fig.~\ref{fig:theta12}, the evolution of $\theta^{}_{12}$ in the five-dimensional UEDM is shown, where the input parameters at $M^{}_Z$ are the same as in the SM and a cutoff scale $\Lambda = 40~{\rm TeV}$ has been chosen to guarantee that a perturbative effective theory is valid. One can observe that the running effect is significant even in such a narrow energy range. It is worthwhile to mention that $\theta^{}_{12}$ increases with respect to an increasing energy scale in both the SM and the UEDM, whereas it decreases in the MSSM.

\newpage 

\noindent {\bf Generic features.} Now, we summarize the generic features of the running of neutrino parameters in the SM, the MSSM, and the UEDM. First, due to small Yukawa couplings of charged leptons in the SM, the evolution of the leptonic mixing angles is insignificant, even in the case of a quasi-degenerate neutrino mass spectrum. The running effects can be remarkably enhanced in the MSSM through a relatively large value of $\tan \beta$, and instead through the number of excited KK modes in the UEDM. Second, among the three leptonic mixing angles, $\theta^{}_{12}$ has the strongest running effect due to an enhancement factor $|\Delta m^2_{31}|/\Delta m^2_{21}$. The running of $\theta^{}_{12}$ in the SM and the UEDM is in the opposite direction to that in the MSSM. However, the actual running behaviour also crucially depends on the choice of the currently unconstrained leptonic CP-violating phases~\cite{Antusch:2003kp}. The running neutrino masses at high-energy scales can be approximately obtained by multiplying a common scaling factor, depending on the evolution of the gauge couplings. Third, the running effects of the leptonic CP-violating phases have been studied in detail in refs~\onlinecite{Antusch:2003kp,Xing:2006sp,Luo:2012ce,Ohlsson:2012pg}, where the evolution of the three CP-violating phases has been found to be entangled. Consequently, a nonzero Dirac CP-violating phase can be radiatively generated even if it is assumed to be zero at a high-energy scale, and vice versa.

Finally, it is worth mentioning that threshold effects may significantly change the running behaviour of different neutrino parameters. However, the accurate description of threshold effects is only possible if the full theory is exactly known.

\section*{Phenomenological implications}
\label{sec:pheno}

\noindent The running of neutrino parameters has important implications for flavour model building, the matter--antimatter asymmetry via the leptogenesis mechanism, and the extra-dimensional models. We now sketch the essential points and refer interested readers to relevant references.

\vspace{5mm}

\noindent {\bf Flavour model building.} In connection with flavour mixing in the quark sector, flavour models are usually built at a high-energy scale, for example, the GUT scale. As for flavour model building, the running effects should be taken into account in general, and for the case of quasi-degenerate neutrino masses in particular. The running effects of mixing parameters can be used to interpret the discrepancy between quark and  lepton flavour mixing~\cite{Mohapatra:2003tw,Haba:2012ar}. As a possible symmetry between quarks and leptons, quark-lepton complementarity relations, such as $\theta^{\rm q}_{12} + \theta^{\rm l}_{12} = 45^\circ$ and $\theta^{\rm q}_{23} + \theta^{\rm l}_{23} = 45^\circ$, where the superscripts specify the mixing angles in the quark and lepton sectors, have been conjectured~\cite{Raidal:2004iw,Minakata:2004xt}. Radiative corrections to these relations have been calculated in the type-I seesaw model~\cite{Schmidt:2006rb}.

To describe the observed lepton mixing pattern, one may impose a discrete flavour symmetry on the generic Lagrangian~\cite{Altarelli:2010gt,King:2013eh}. As discussed in the previous section, a bi-maximal mixing pattern (that is, $\theta^{}_{12} = \theta^{}_{23} = 45^\circ$ and $\theta^{}_{13} = 0$) at $\Lambda^{}_{\rm GUT}$ turns out to be compatible with current neutrino oscillation data if running effects are taken into account~\cite{Antusch:2002hy}. In addition to bi-maximal mixing~\cite{Barger:1998ta}, tri-bimaximal~\cite{Harrison:2002er,Harrison:2002kp,Xing:2002sw}, democratic~\cite{Fritzsch:1995dj}, and tetra-maximal mixing~\cite{Xing:2008ie} patterns have been proposed to describe lepton flavour mixing, and their radiative corrections have also been examined~\cite{Miura:2003if,Luo:2005fc,Dighe:2006sr,Boudjemaa:2008jf,Xing:2000ea,Mei:2005gp,Zhang:2011aw}.

\vspace{5mm}

\noindent {\bf Matter--antimatter asymmetry.} It remains an unanswered question why our visible world is made of matter rather than antimatter. From cosmological observations, the ratio between baryon number density and photon number density $\eta^{}_{\rm b} = (6.19\pm 0.15) \times 10^{-10}$ has been precisely determined~\cite{PDG:2012}. One of the most attractive mechanisms for a dynamic generation of baryon asymmetry is leptogenesis~\cite{Fukugita:1986hr}, which works perfectly in various seesaw models for neutrino mass generation.

Take the type-I seesaw model for example, where three heavy right-handed neutrinos are introduced. In the early universe, when the temperature is as high as the masses of heavy neutrinos, they can be thermally produced and decay into the SM particles, mainly lepton and Higgs doublets. If the neutrino Yukawa couplings are complex, heavy neutrinos decay into leptons and anti-leptons in different ways. When the universe cools down, CP-violating decays go out of thermal equilibrium and a lepton asymmetry can be generated, which will be further converted into a baryon asymmetry.

The final baryon asymmetry $\eta^{}_{\rm b} \approx 0.96\times 10^{-2} \varepsilon^{}_1 \kappa^{}_{\rm f}$ depends on the CP asymmetry $\varepsilon^{}_1$ from the decays of the lightest heavy neutrino, and the efficiency factor $\kappa^{}_{\rm f}$ from the solution to a set of Boltzmann equations~\cite{Buchmuller:2004nz}. Moreover, the maximal value of $\varepsilon^{}_1$ can be derived
\begin{equation}
\varepsilon^{\rm max}_1 \approx  \frac{3 M^{}_1 |\Delta m^2_{31}|}{16 \pi^2 \langle H \rangle^2 m} \; ,
\label{eq:CP}
\end{equation}
where $m$ denotes the mass of heaviest ordinary neutrino~\cite{Buchmuller:2003gz}. Now, it is evident that the running of neutrino masses from the low-energy scale to $M^{}_1$ (that is, the mass of $\nu^{}_{1{\rm R}}$) should be taken into account~\cite{Antusch:2005gp,Cooper:2011rh}. As the evolution of neutrino masses can be described by a common scaling factor and they become larger at a higher-energy scale, the maximum of the CP asymmetry scales upwards as neutrino masses. However, larger values of neutrino masses at $M^{}_1$ imply larger Yukawa couplings, which enhance the washout of the lepton asymmetry, and thus reduce $\kappa^{}_{\rm f}$. The outcome from the competition between the enhancement of $\varepsilon^{}_1$ and the reduction of $\kappa^{}_{\rm f}$ depends on the neutrino mass spectrum, and also on the value of $\tan \beta$ in the MSSM~\cite{Antusch:2003kp}. See ref.~\onlinecite{Hambye:2012fh}, for a review on the recent development of leptogenesis in seesaw models.

\vspace{5mm}

\noindent {\bf Bounds on extra dimensions.} A general feature of quantum field theories with extra spatial dimensions is that they are non-renormalizable~\cite{Dienes:1998vg}, since there exist infinite towers of KK states appearing in the loops of quantum processes. As pointed out in ref.~\onlinecite{Dienes:1998vg}, the higher-dimensional theories could preserve renormalizability
if they are truncated at a certain energy scale $\Lambda$ (see Fig.~\ref{fig:theta12}), below which only a finite number of KK modes is present. In the UEDM, $\Lambda$ is usually taken to be the energy scale where the gauge couplings become non-perturbative~\cite{Hooper:2007qk}, but it could also be related to a unification scale for the gauge couplings~\cite{Dienes:1998vg}.

The recent discovery of a Higgs particle with $M^{}_H = 126~{\rm GeV}$ leads to a reconsideration of the stability of the SM vacuum~\cite{Xing:2011aa,Buttazzo:2013uya}. The instability is essentially induced by the fact that the Higgs self-coupling constant $\lambda$ runs into a negative value at a high-energy scale. Since the model parameters have a power-law running in the UEDM, in contrast to a logarithmic running in the ordinary four-dimensional theories, the requirement of vacuum stability will place a restrictive bound on the cutoff scale $\Lambda$ and the radius of extra dimensions $R$. It has been found that $\Lambda R < 5$ for $R^{-1} = 1~{\rm TeV}$ in the five-dimensional UEDM~\cite{Blennow:2011tb}, while this bound becomes more stringent $\Lambda R < 2.5$ in the six-dimensional UEDM~\cite{Ohlsson:2012hi}, which can be translated into the maximal number of KK modes being five and two, respectively. As a consequence, the running of neutrino parameters in these models will be limited to a narrow energy range.

\section*{Outlook}
\label{sec:outlook}

\noindent Our knowledge about neutrinos has been greatly extended in the past decades, especially due to a number of elegant neutrino oscillation experiments. As for the leptonic mixing parameters, we are entering into the era of precision measurements of three leptonic mixing angles and two neutrino mass-squared differences. The determination of the neutrino mass hierarchy and the discovery of leptonic CP violation are now the primary goals of the ongoing and upcoming neutrino oscillation experiments. On the other hand, the tritium beta decay and neutrinoless double-beta decay experiments, together with cosmological observations, will probe the absolute scale of neutrino masses. Whether or not neutrinos are their own antiparticles will also be clarified if neutrinoless double-beta decay is observed. Therefore, we will obtain more information about the neutrino parameters at the low-energy scale.

However, the origin of neutrino masses and lepton flavour mixing remains a
big puzzle in particle physics. In this review article, we have elaborated
on the evolution of neutrino parameters from the low-energy scale to a
superhigh-energy scale, where new physics may appear and take the responsibility for generating neutrino masses. The running effects of neutrino parameters can be very significant and should be taken into account in searching for a true theory of neutrino masses and lepton flavour mixing. On the other hand, the successful applications of renormalization group running in neutrino physics, and more generally in elementary particle physics and condensed matter physics, will demonstrate the deep connection between different branches of physical sciences and the amazing power of quantum field theories in describing Nature.

In the foreseeable future, with direct searches at colliders and precision measurements of quark and lepton flavour mixing parameters, we hope that all observations will finally converge into hints for new physics beyond the SM and lead us to a complete theory of fermion masses, flavour mixing, and CP violation.

\begin{acknowledgments}
\noindent The authors would like to thank Mattias Blennow and He Zhang for useful discussions and comments. This work was supported by the Swedish Research Council (Vetenskapsr{\aa}det), contract no. 621-2011-3985.
\end{acknowledgments}

\section*{Additional information}
\noindent {\bf Competing financial interests:} The authors declear no competing financial interests.


\end{document}